\documentclass[twocolumn,reprint,amsmath,amssymb,aps,pra]{revtex4-1}

\usepackage{graphicx}
\usepackage{bm}
\usepackage{color}
\usepackage{natbib}
\usepackage{epstopdf}
\usepackage{amsmath}
\usepackage{tabularx}
\usepackage{float}

\begin{document}

\title{Resonant excitation of acoustic waves in one-dimensional exciton-polariton systems}

\author{A. V. Yulin$^{1}$}
\author{V. K. Kozin$^{2,1}$}
\author{A. V. Nalitov$^{2,1}$}
\author{I. A. Shelykh$^{2,1}$}

\affiliation{$^1$ITMO University, Kronverkskiy prospekt 49, Saint Petersburg 197101, Russia}

\affiliation{$^2$Science Institute,
University of Iceland, Dunhagi 3, IS-107, Reykjavik, Iceland}

\begin{abstract}
We analyze the interaction of exciton-polariton condensates in a 1D semiconductor microcavity with acoustic phonons. We consider the case of a coherently pumped condensate and demonstrate that upon passing of a certain threshold the parametric instability in the system leads to the generation of a coherent acoustic wave and additional polariton harmonics. The process is strongly affected by exciton-exciton interactions which lead to the appearance of the effects of the bistability and hysteresis in the system.
\end{abstract}

%\pacs{Valid PACS appear here}
%use \showpacs to show PACS
\maketitle

%\tableofcontents

\section{Introduction}

Many-body interactions can dramatically modify the properties of condensed matter systems. The textbook examples are 1D gas of interacting fermions where instead of a normal Fermi liquid Tomonaga-Luttinger liquid with excitations of a bosonic type appears \cite{Guan2013} and 2D electron gas in a strong magnetic field where in the regime of the fractional Quantum Hall effect incompressible Laughlin liquid with excitations possessing fractional charge and anyonic statistics forms \cite{Murthy2003}. If quasiparticles of different type interact, new elementary excitations of a hybrid nature can appear if characteristic interaction energy exceeds all characteristic broadenings in the system and strong coupling regime is achieved. In particular, coupling between phonons in ionic crystals and photons leads to the formation of phonon polaritons \cite{Huang1951}, and coupling between excitons in semiconduсtor quantum wells and confined electromagnetic mode of a planar resonator gives rise to exciton-polaritons \cite{KavokinBook}. The latter possess a set of remarkable properties which make them  unique laboratory for the study of quantum collective phenomena at high temperatures. 

Having extremely small effective mass ($10^{-4}-10^{-5}$ of the effective mass of free electrons), obeying bosonic statistics and efficiently interacting with each other, polaritons can undergo a transition into a collective phase usually referred to as polariton BEC \cite{Kasprzak2006}
accompanied by the onset of superfluidity \cite{Amo2009,Amo2009a}. Polaritons can also interact with other types of the quasiparticles present in the system. Polariton-electron interaction can lead to the onset of the polariton lasing in the system \cite{Malpuech2003} or even transition to a superconducting regime where analogs of the Cooper pairs are formed due to the exchange of the bogolons, virtual excitations of a polariton condensate \cite{Laussy2010}. Interaction with incoherent reservoir of thermal phonons can lead to the thermalization of a polariton system \cite{Haug2005} necessary for the formation of a polariton BEC \cite{Malpuech2003a}. On the other hand, interaction with coherent acoustic wave \cite{Jusserand2007} can lead to such effects as formation of hybrid exciton-acoustopolaritons \cite{Vishnevsky2011}, the possibility of the observation of the polariton Bloch oscillations \cite{Flayac2011}, and polariton condensation in the dynamic polariton lattices \cite{Krizhanovskii2010,Krizhanovskii2012}.

In the present paper we consider further the interaction between 1D condensate of exciton-polaritons and a coherent phonon field, possessing the same dimensionality. From an experimental point of view, this setup can be realized as a ring-shaped polariton microcavity~\cite{PhysRevB.98.125115} on a substrate that prevents leakage of the generated acoustic waves from the microcavity to the substrate. The radius of the microcavity should be big enough to neglect the discreteness of the spectrum in the ring. We use the model of the interaction between excitons and phonons written in terms of the classical fields developed before\cite{Vishnevsky2011,Ivanov2001}, and consider the case of the coherent optical pump of a polariton mode. We demonstrate that after passing of a certain threshold of pumping strength the parametric instability occurs and the system generates additional polariton harmonics together with a coherent acoustic wave. We also demonstrate that polariton-polariton interactions crucially affect this process leading to the development of additional instabilities related to the effects of the bistability and hysteresis.

\section{Analytical model}
Let us start from the coupled equations describing coherent polariton-phonon interactions in 1D  \cite{Vishnevsky2011}:
\begin{align}\label{eq:pol_main}
    \mathrm{i \hbar} \partial_t \Psi =& \left[ -{\hbar^2 \partial^2_x \over 2 m} + \alpha|\Psi|^2 +g \Phi - i \hbar \gamma \right] \Psi + P(x,t),\\
\label{eq:sound_main}
    \partial^2_t \Phi =& {1 \over \rho} \partial_x^2 \left[ Y\Phi + g \vert \Psi \vert^2 \right] - \Gamma  \partial_t \Phi,
\end{align}
where $\Psi$ and $\Phi$ are the polariton and phonon scalar fields, $m$ is the polariton effective mass, $\alpha$ and $g$ are the polariton-polariton and polariton-phonon interaction constants, $\gamma$ and $\Gamma$ are the polariton and phonon decay rates, $Y$ and $\rho$ are the crystal Young's modulus and density, which are related to the speed of sound $c=\sqrt{Y/\rho}$.
 
We rewrite the Eqs. (\ref{eq:pol_main},\ref{eq:sound_main}) in dimensionless form by redefining $t\rightarrow t/t_0$, $x \rightarrow x / x_0$, $\gamma \rightarrow \gamma t_0$, $\Gamma \rightarrow \Gamma t_0$, $\Psi \rightarrow \Psi \sqrt{x_0}$, $\Phi \rightarrow \Phi \sqrt{\rho c x_0^2/\hbar}$, $\alpha \rightarrow \alpha / \hbar c$, $g \rightarrow g / \sqrt{\hbar \rho c^3}$, $P \rightarrow - \mathrm{i} t_0 \sqrt{x_0} P / \hbar$, where $t_0 = \hbar /(m c^2)$, $x_0 = \hbar/(m c)$:

\begin{align} \label{eq_GPE1}
\partial_t \Psi =& \left [ \frac{i}{2}\partial_x^2 - i g\Phi - i \alpha \vert \Psi \vert^2 - \gamma \right] \Psi + P, \\ \label{eq_sound1}
\partial_t^2 \Phi =& \partial_x^2 \left[ \Phi + g |\Psi|^2 \right] - \Gamma  \Phi.
\end{align}

The polariton-phonon coupling terms with prefactor $g$ are in focus of this work.
In the following we will first consider truncated model with omitted polariton-polariton interaction, decay, and pumping terms.
This simplified model allows straightforward solution and easy to follow derivations that provide insights into polariton condensate instabilities stemming from coherent acousto-polaritonic interaction.
The full model is treated in the rest of the section and addresses the interplay of the optical bistabilities, generated by polariton-polariton interactions, and acoustic instabilities of polariton condensate, as well as effects of the losses and continuous coherent pumping.

\subsection{Simplified model}

Let us first neglect polariton-polariton interactions, external pumping and decay, focusing on stability of an initially prepared single mode state to nonlinear generation of harmonics. The case when the polariton-polariton interaction can be neglected corresponds, for instance, to TMDC materials with a large oscillator strength so it has a direct physical meaning but this section will also be a scene for the further consideration.  
The equations (\ref{eq_GPE1},\ref{eq_sound1}) then transform to:
\begin{align} \label{eq_GPE2}
\partial_t \Psi = \frac{i}{2}\partial_x^2 \Psi - i g\Phi \Psi, \\ \label{eq_sound2}
\partial_t^2 \Phi-\partial_x^2 \Phi= g \partial_x^2 |\Psi|^2 
\end{align}
The system (\ref{eq_GPE2},\ref{eq_sound2}) has a trivial stationary solution of the form 
\begin{equation} \label{eq_trivial}
\Psi = \psi_0 \exp(-i\Omega t +i\kappa x), \; \Phi = 0.
\end{equation}
Without loss of generality $\psi_0$ can be assumed to be real.
To study the stability of the trivial solution \eqref{eq_trivial} we introduce small variations $\Psi=(\psi_0 + \psi)\exp(-i\Omega t +i\kappa x)$, $\Phi=\phi$.
The equations on the corrections $\psi$ and $\phi$ read:
\begin{align}
\partial_t \psi =& \frac{i}{2}\partial_x^2 \psi - \kappa \partial_x \psi - i G \phi,  \\
\partial_t^2 \phi -&\partial_x^2 \phi = G \partial_x^2 (\psi+\psi^{*}),
\end{align}
where we defined $G = g\psi_0$ for convenience.
Let us represent the complex field $\psi$ as $\psi=\psi_1+i\psi_2$ where $\psi_1$ and $\psi_2$ are real.
Now we can write down linear equations for $\psi_1$ and $\psi_2$:
\begin{align}
\partial_t \psi_1=&-\frac{1}{2}\partial_x^2 \psi_2 - \kappa \partial_x \psi_1, \label{eq_real}\\ 
\partial_t \psi_2=&\frac{1}{2}\partial_x^2 \psi_1- \kappa \partial_x \psi_2 - G \phi, \label{eq_imag}\\ 
\partial_t^2 \phi-\partial_x^2 \phi=& 2G \partial_x^2 \psi_1. \label{eq_sound3}
\end{align}
Searching for the solution in the form $\psi, \phi \propto \exp(-i\omega t +ikx)$ we arrive to a set of algebraic equations
\begin{align}
i(\kappa k-\omega) \psi_1=&-\frac{k^2}{2} \psi_2, \label{eq_alg_real}\\ 
i(\kappa k -\omega) \psi_2=&-\frac{k^2}{2} \psi_1 - G \phi, \label{eq_alg_imag}\\ 
(\omega^2-k^2) \phi =& 2G k^2 \psi_1 \label{eq_alg_sound},
\end{align}
from where we obtain the equation on the energy dispersion $\omega(k)$:
\begin{align}
\left[ \omega - \kappa k + \frac{k^2}{2}  \right]
\left[ \omega - \kappa k - \frac{k^2}{2}  \right] %\times \nonumber \\
\left(\omega-k\right)\left(\omega+k\right)=G^2k^4.
\end{align}

One may note that in the absence of the coupling term $G=0$ the equation yields the separated dispersion characteristics of the polaritons ($\omega_{p1} = \kappa k - k^2/2$, $\omega_{p2} = \kappa k + k^2/2$) and the sound ($\omega = \pm k$).
In the following we will assume the coupling to be weak and distinguish two important regimes.

The first one corresponds to "supersonic polaritons" and is reached when the following condition is satisfied:
$$\partial_k \omega_{p1,2}|_{k=0}>1.$$
For positive $k$ the resonant interaction between phonons and polaritons takes place at
$$ \kappa k_1 - \frac{k_1^2}{2}=k_1$$
(the first resonance that will be referred to as "r1") and 
$$ \kappa k_2 + \frac{k_2^2}{2}=-k_2$$
(the second resonance that will be referred to as "r2"). Of course there are symmetric resonances at negative $k$ with identical properties.

First we consider resonance "r1" at $k_1=2(\kappa-1)$.
We introduce the detuning of the wave vector from the resonance $q=k-k_1$ and the detuning of the frequency $\delta=\omega-\omega_{p1}(k_1)$.
Then, keeping only the leading terms, we obtain the dispersion relation equation expressed in $\delta$ and $q$:
\begin{equation}
(\delta-q)(\delta - v_1 q)=-4G^2(\kappa-1),
\end{equation}
where $v_1=\partial_k \omega_{p1}|_{k=k_1}=1-\kappa$, this is nothing else but the group velocity of the polaritons at the resonant point. 

The equation for $\delta$ is quadratic and has the solutions
\begin{equation}
\delta = \frac{q(1+v_1)\pm \sqrt{(1-v_1)^2q^2-4\rho_1^2}}{2},
\end{equation}
where $\rho_1^2=4G^2(\kappa-1)$.
One may note that for 
\begin{equation}
|q|<\frac{2\rho_1}{|1-v_1|}
\end{equation}
one of the two eigenmodes has a positive imaginary part of energy, which indicates an instability resulting in coherent generation of phonons with simultaneous polariton scattering into a new mode.

In a similar fashion we consider the resonance "r2" at $k_2=\kappa+1$.
In this case the dispersion equation for $\delta$ and $q$ reads: 
\begin{equation}
(\delta+q)(\delta - v_2 q)=4G^2(1+\kappa),
\end{equation}
where $v_2=\partial_k \omega_{p1}|_{k=k_2}=\kappa-k_2$.
The solution now reads:
\begin{equation}
\delta = \frac{-q(1-v_2)\pm \sqrt{(1+v_2)^2q^2+4\rho_2^2}}{2},
\end{equation}
where $\rho_2^2=4G^2(\kappa+1)$.
Note that in contrast to the "r1" resonance in this case $\delta$ is real, therefore "r2" resonance does not lead to an instability.

In the opposite regime of subsonic polaritons, the derivations for the "r2" resonance are identical.
In contrast, the resonance "r1", which is at $$\tilde k_1=1-\kappa,$$ in this case, is generated by a different crossing of unperturbed dispersions.
The equation on $\delta$ and $q$ is correspondingly modified:
\begin{equation}
(\delta-q)(\delta - \tilde v_1 q)=4G^2(1-\kappa),
\end{equation}
where $\tilde v_1=\partial_k \omega_{p2}|_{k=\tilde k_1}=\kappa+\tilde k_1$.
From its solution
\begin{equation}
\delta = \frac{q(1+\tilde v_1)\pm \sqrt{(1-\tilde v_1)^2q^2+4\tilde \rho_1^2}}{2},
`\end{equation}
with $\tilde \rho_1^2=4G^2(1-\kappa)$, one may see that this resonance also does not result in an instability.

\subsection{Extended model}

We now include weak polariton-polariton interactions as well as external monochromatic optical pumping and decay of both polaritons and phonons.
The coupled dimensionless equations on the polariton and phonon order parameters read:
\begin{align}
\partial_t \Psi & = \left[ \frac{i}{2}\partial_x^2 - i \alpha|\Psi|^2  - i g \Phi - \gamma \right] \Psi + P e^{i\kappa x -i\delta t},
\label{master_equation1} \\
\partial_t^2 \Phi & = \partial_x^2 \left[ \Phi + g |\Psi|^2 \right] - \Gamma \partial_t \Phi. \label{master_equation2}
\end{align}

By the transformation $\Psi \rightarrow \Psi \exp(i\kappa x -i\delta t) $ we switch to a rotating frame for convenience:
\begin{equation}
\partial_t \Psi = \left[ \frac{i}{2}\partial_x^2 - \kappa \partial_x + i \mu - i \alpha|\Psi|^2  - i g \Phi - \gamma \right] \Psi +P,
\label{master_equation1_}
\end{equation}
where $\mu = \delta - \kappa^2/2$ is the detuning of the pumping frequency from the polaritonic resonance.

In the absence of coherent acoustic component the pump generates spatially uniform state, corresponding to the trivial solution of the truncated model \eqref{eq_trivial}.
Depending on the detuning $\mu$ and polariton losses $\gamma$ this system may support bistability \cite{Vishnevsky2011}.
It stems from the polariton-polariton repulsion and follows from the nonlinear relation of the pump with the population $N_0 = \vert \Psi \vert^2$ of the polariton state:
\begin{eqnarray}
|P|=\sqrt{ \left[ (\mu+\alpha N_0)^2+\gamma^2 \right] N_0}. \label{bif_diag_0}
\end{eqnarray}

The spatially uniform state can be unstable with respect to either polariton-polariton or polariton-phonon interactions.
The instability due to polariton-polariton interaction is well studied \cite{Gippius2007}.
For the typical case of defocusing nonlinearity, corresponding to repulsive interactions, the upper branch of the bifurcation diagram is stable, the intermediate is unstable, and the lower branch is partially unstable with the critical point of instability depending on the losses.
Here, on the contrary, we address the instability stemming from polariton-phonon interactions. 

Similarly to the previous section, we search for the solution in the form of the a perturbed spatially uniform wave function $\Psi=\psi_0+\psi$ with $\psi_0 = \sqrt{N_0}$ chosen real without loss of generality. As there is a one-to-one correspondence between $\Psi$ and $\psi$, we will be referring to the obtained dispersion for $\psi$ as a polariton dispersion.
To investigate the polariton-phonon instability it is necessary, of course, to account for the interaction with phonons. The phonon field $\Phi = \phi$ produced by the perturbation has the same order of smallness, thus we may linearize Eqs. (\ref{master_equation1_},\ref{master_equation2}):
\begin{align}
\label{inst_equation1}
\partial_t \psi =& \left[ \frac{i}{2}\partial_x^2 - \kappa \partial_x + i\mu - 2 i A - \gamma \right] \psi -i A \psi^* -i G \phi, \\
\partial_t^2 \phi =& \partial_x^2 \left[ \phi +  G \left(\psi +\psi^{*} \right) \right] - \Gamma \partial_t \phi, \label{inst_equation2}
\end{align}
where $G = g \psi_0$ and $A = \alpha N_0$.

Eqs. \eqref{inst_equation1}-\eqref{inst_equation2} can be expressed in a form of a matrix operator equation: 
\begin{eqnarray}
\partial_t Y = \hat L Y,
 \label{inst_equation_vec}
\end{eqnarray}
where $Y = \left( \mathrm{Re} \lbrace \psi \rbrace, \mathrm{Im} \lbrace \psi \rbrace, \phi, \partial_t \phi \right)^T $:
\begin{eqnarray}
\hat L =\left( \begin{array}{cccc}
  -\gamma - \kappa \partial_x   &  -\mu -\frac{1}{2}\partial_x^2 - A & 0 & 0 \\
    \mu +\frac{1}{2}\partial_x^2 + 3 A  & -\gamma - \kappa \partial_x & - G & 0 \\
   0 & 0 & 0 & 1 \\
   2 G \partial_x^2 & 0 & \partial_x^2 & -\Gamma
\end{array}     \right).
\label{inst_operator}
\end{eqnarray}

Finally, we can find the plane wave eigenmodes $Y \propto \exp(ikx-i\omega t)$ and the corresponding eigenfrequencies.
After some algebra we obtain the equation on the dispersion relation $\omega(k)$ for the waves propagating in the system in the convenient form
\begin{align}
\left( \omega -\kappa k +D(k) +i\gamma \right)\left( \omega -\kappa k -D(k) +i\gamma \right) \times& \nonumber \\
\left( \omega +\sqrt{k^2-\frac{\Gamma}{4}+\frac{i\Gamma}{2}}\right)\left( \omega -\sqrt{k^2-\frac{\Gamma}{4}+\frac{i\Gamma}{2}}\right) &= \nonumber \\
G^2 k^2 \left[ k^2-2(\mu+A) \right],
 \label{disp_equation}
\end{align}
where 
\begin{eqnarray}
D(k) = \sqrt{ \left( \mu +3 A -\frac{k^2}{2}\right) \left(\mu + A -\frac{k^2}{2} \right)}.
\end{eqnarray}

Similarly to the previous section, in the absence of the polariton-phonon interaction term $G=0$ Eq. \eqref{disp_equation} may be analytically solved.
Each of the brackets in the left hand side of the equation generates a dispersion characteristics of either polaritons or phonons.
The first two brackets correspond to polaritons propagating to the right and to the left respectively.
Note that the polariton dispersions in this case accounts for the polariton-polariton interaction and thus inherits the Bogoliubov excitation properties.
The second pair of brackets in turn corresponds to the acoustic waves.

In the general case $G \neq 0$ analytic solution of Eq. \eqref{disp_equation} is problematic, however, it can readily be solved numerically, or, as in the previous section, can be analyzed with the perturbation theory in the case of weak interaction and small losses.
Obviously, the polariton-phonon interactions are mostly pronounced at the crossings of the unperturbed polariton and acoustic energy dispersion curves.
In the vicinity of these points the solution of Eq. \eqref{disp_equation} can be sought in the form $\omega=\omega_r+\Delta$ and $k=k_r+q$ where $\omega_r$ and $k_r$ are the frequency and the wave vector of the resonance given be the crossing of the dispersion characteristics of polaritons and the phonons.
The polariton-phonon coupling and the losses in the polariton and the phonon subsystems are then considered as perturbations yielding a quadratic equation on the resonance shift $\Delta$.

Let us consider in detail the crossing of the acoustic branch $\omega=k$ and polariton branch $\omega = \kappa k - D$. Introducing the group velocities as $v_p=\partial_k \omega_p=\partial_k (\kappa k -D)$ and $v_s=\partial_k \omega_s=1$ we can simplify the dispersion relation in the vicinity of the resonance.
Keeping only the leading terms we obtain the reduced equation:
\begin{equation}
\left( \Delta - v_p q + i\gamma \right) \left (\Delta -v_s q +\frac{i\Gamma}{2} \right) = 
R, \label{disp_eq_smpl}
\end{equation}
where
$$
R = \frac{G^2 k_r^2 \left[ k_r^2-2(\mu+A) \right]}{4\omega_r(\omega_r-\kappa k)}
$$
Its solution reads:
\begin{align}
\Delta = &{v_p + v_s \over 2}q - {i \over 2}\left( \gamma + {\Gamma \over 2} \right) \pm \nonumber \\
 &\sqrt{ \left[ {v_p+v_s \over 2}q-{i\over 2} \left(\gamma+\frac{\Gamma}{2}\right) \right]^2 + R },
 \label{disp_solution}
\end{align}
We omit here the similar expressions for the resonance shifts for other dispersion crossing points for the sake of brevity.

\subsection{Two-mode approximation}

From the renormalized positions of the resonances we develop a theory allowing to find the stationary amplitudes of the directly and parametrically pumped polariton states.
The main assumption of the theory is that only one spatially uniform polariton plane wave mode is generated by the acousto-polariton coupling.
Although cascade generation of a multi-mode polariton lasing state is possible at high pumping powers, as further discussed, the initial instability populates a single dominant polariton mode and a corresponding acoustic mode, which proves our assumption valid for pumping powers slightly above the first instability threshold.
The condition on the choice of the parametrically generated mode is governed by the resonance: the mode at the energy dispersion crossing point has the largest growth rate and thus it is reasonable to assume that this mode wins the mode competition.
It allows us to search for the solution in the form $\Psi=a_1 + a_2 \exp(i k_r x- i\omega_r t)$ where $a_1$ is the amplitude of the directly excited mode, $a_2$ is the amplitude of the parametrically excited mode, $\omega_r$ and $k_r$ are the resonant frequency and wave vector.
Here we also assume that the amplitude of the parametrically excited mode is weak $\vert a_2 \vert \ll \vert a_1 \vert$, which is also valid in the vicinity of the instability threshold.

From Eq. \eqref{master_equation2} we have for the acoustic mode:
\begin{equation} \label{eq_acoustic1}
\Phi = -\frac{ gk_r^2 a_1 a_2^{*}\exp(-i k_r x+ i\omega_r t)  }{ k_r^2-\omega_r^2 + i\Gamma \omega_r   }+c.c.
\end{equation}
As the resonant condition $\omega_r=k_r$ is satisfied, the expression \eqref{eq_acoustic1} reduces to
\begin{eqnarray}
\Phi = i \frac{ g k_r}{ \Gamma} a_1 a_2^{*}\exp(-i k_r x+ i\omega_r t) + c.c.
\label{ampl_ac}
\end{eqnarray}

Substituting the expression \eqref{ampl_ac} into Eq. \eqref{master_equation1} and keeping the terms proportional to $\exp(i k_r x- i\omega_r t)$ we obtain
\begin{equation}
\gamma a_2 = \frac{ g^2 k_r }{ \Gamma  } \vert a_1 \vert^2 a_2,
\end{equation}
from where we express the population of the directly excited mode $N_1 = \vert a_1 \vert^2$ above its instability threshold ($a_2 \neq 0$)
\begin{equation}
N_1 =\frac{\gamma \Gamma}{g^2 k_r},
\label{a1}
\end{equation}
from where we see that for an experimental observation one should decrease the loses as much as possible. 
The main mode population thus remains at the level allowing the parametric gain to exactly compensate the losses of the secondary mode.
Note that the effect of Kerr nonlinearity (polariton-polariton interaction) is accounted for: in the leading term of approximation it only enters the expression \eqref{a1} for $a_1$ via the resonant wavevector dependence on the amplitude of the directly excited state.

The position of the perturbed resonance $\omega_r = k_r$ may now be self-consistently derived by substituting Eq. \eqref{a1} for the directly pumped mode occupation into Eq. \eqref{disp_solution} for the resonance shift.

Collecting the time independent terms of Eq. \eqref{master_equation1} we obtain an equation on the secondary mode occupation $N_2 = \vert a_2 \vert^2$:
\begin{equation}
    \left[ i \mu - i \alpha \left( N_1 + 2 N_2 \right) - {g^2 k_r \over \Gamma} N_2 - \gamma \right] a_1 + P = 0,
\end{equation}
which may be converted into a quadratic equation
\begin{equation}
\left[ \gamma + \frac{g^2 k_r}{\Gamma}N_2 \right]^2 + \left[ \mu - \alpha \left( N_1+2N_2 \right) \right]^2 = {P^2 \over N_1},
\end{equation}
from where we derive for $N_2$:
\begin{align} \label{a2}
& {N_2 \over N_1} = {2 A_1 \left( \mu - A_1 \right) - \gamma^2 \over 4 A_1^2 + \gamma^2} \pm  \\
& \sqrt{\left[ 2 A_1 \left( \mu - A_1 \right) - \gamma^2 \over 4 A_1^2 + \gamma^2 \right]^2 + {P^2 / N_1 - \left( \mu - A_1 \right)^2 - \gamma^2 \over 4 A_1^2 + \gamma^2} } \nonumber
\end{align}
where $A_1 = \alpha N_1$.
If the polariton-polariton interaction is neglected ($A_1 = 0$) expression (\ref{a2}) is substantially simplified:
\begin{eqnarray}
N_2=\frac{N_1}{\gamma}\sqrt{{P^2 \over N_1} + \mu^2  } - N_1.
\label{a2+noKerr}
\end{eqnarray}

\section{Comparison of the analytical  results against direct numerical simulations}

\subsection{Properties of linear excitations in the absence of nonlinear polariton-polariton interactions}

Now we can use the developed theory to analyze the behaviour of the system, starting from the case without polariton-polariton interactions $\alpha=0$ and pump having the wavevector large enough so that the velocity of the polaritons excited by the pump is higher than the speed of sound. 

\begin{figure}[h]
\includegraphics[width=\linewidth]{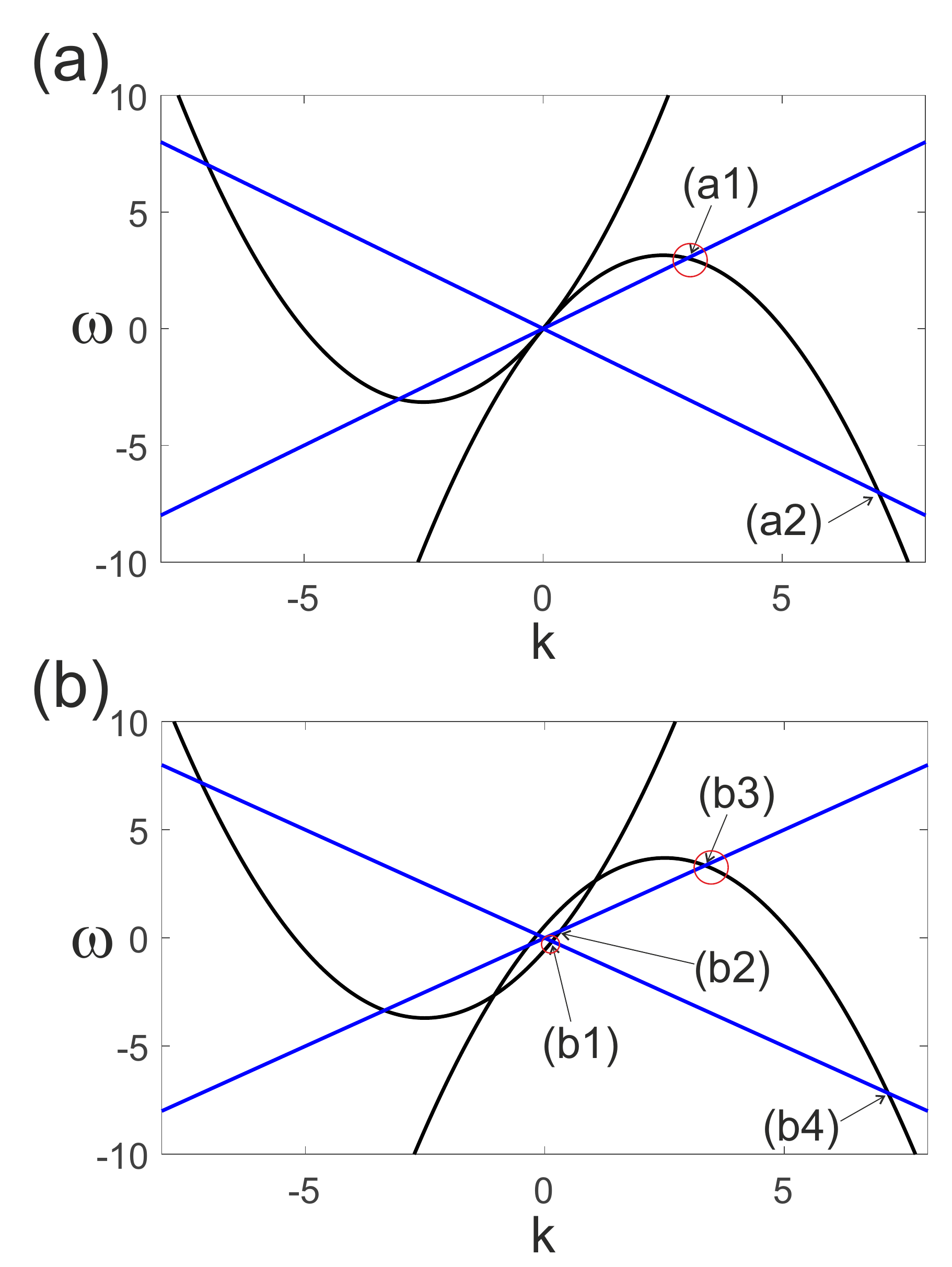}
\caption{The dispersion characteristics of the weak excitations in the case of the pump with a zero (a) and non-zero detuning (b).
The blue lines correspond to the acoustic waves and black lines - to the polariton components. The wavevector of the pump is $\kappa=2.5084$, the frequency of the pump is $\delta=3.146$ and $\mu=3.7$ for the cases with a zero and non-zero detuning correspondingly. The red circles mark the crossings that produces the instabilities. }
\label{fig:1}
\end{figure}

The solution of the dispersion equation is shown in Fig.~\ref{fig:1} for the parameters of the pump $\partial_k \omega_p(k)|_{\kappa}>1$.
The polariton-polariton interactions and the phonons and the losses are neglected. In this plot we also neglected the interaction between the polaritons and the phonons, this interaction is considered below by perturbation theory. 

\begin{figure}
\includegraphics[width=\linewidth]{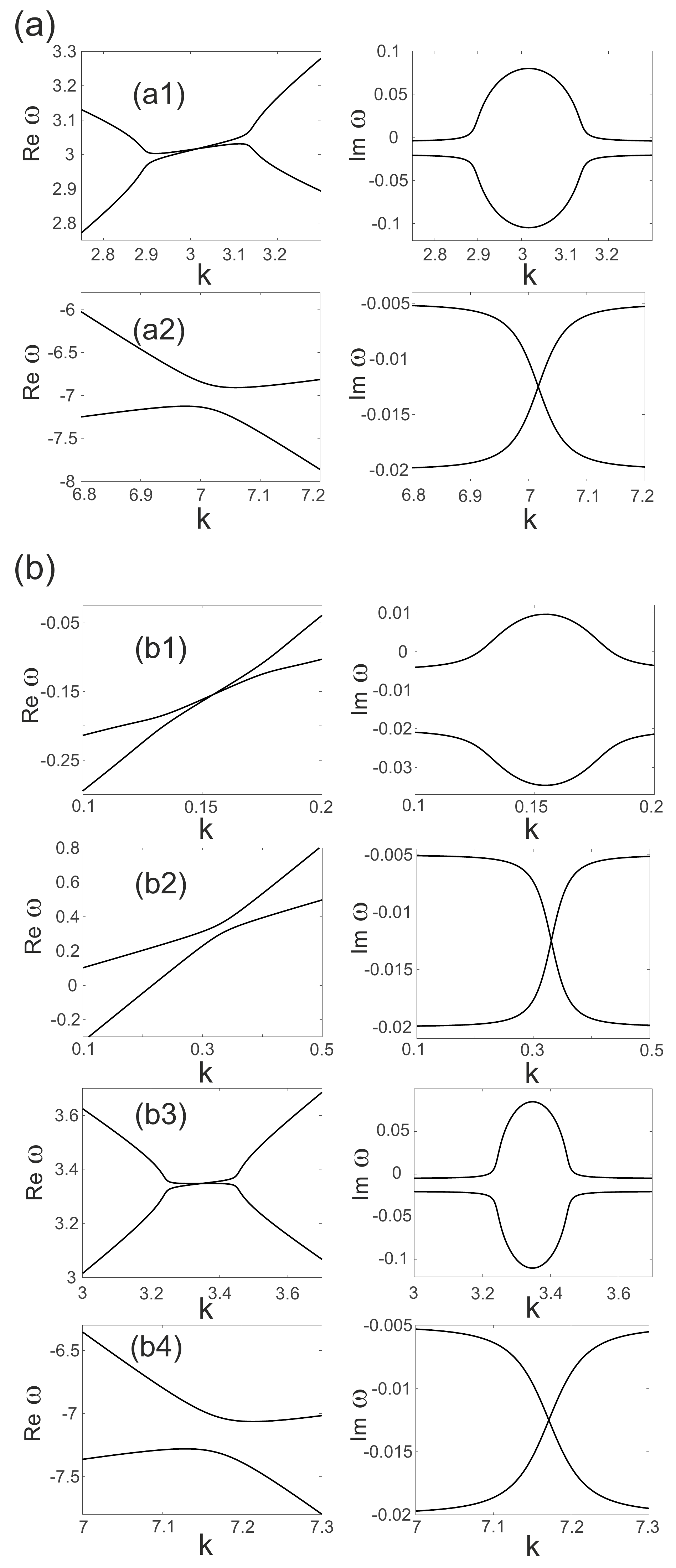}
\caption{(a) The dependencies of the real (left column) and imaginary (right column) parts of the frequencies of the weak excitations as functions of their wavevectors. The row marked as (a1) corresponds to the crossing marked as (a1) in Fig.~\ref{fig:1}(a), the row marked as (a2) corresponds to the crossing marked as (a2) in Fig.~\ref{fig:1}(a).  
(b)The same as in (a) but for the pump with a detuning $\mu=3.7$. }
\label{fig:2}
\end{figure}

The dispersion characteristics calculated perturbatively by formula (\ref{disp_solution}) in the vicinity of the crossing point are shown in Fig.~\ref{fig:2} for the case of the resonant pump and weak losses $\gamma=0.02$ and $\Gamma=0.01$, the acousto-polariton interaction constant is $g=0.075$.  The population of the initial polariton state is $N_0=1.0$ which at the given parameters is provided by the pump of the amplitude $P=0.02$. 

It is seen that there is a region of wave vectors where imaginary part of the frequency becomes positive, which means that the perturbations with these wave vectors grow exponentially.
The second resonance results in the splitting of the dispersion characteristics and to the formation of a hybrid mode.

To check the analytic results we performed direct numerical simulation.
First we set the polariton-phonon coupling to zero $g=0$ and performed numerical simulation for the time much longer than the relaxation time.
By this we provide that a stationary polariton state is formed.
Then we perturbed the state by weak noise in both the polariton and the acoustic components and introduced final coupling $g=0.075$ between the polaritons and the phonons.
Then we continued numerical simulations.

To characterize the dynamics of the field it is convenient to introduce the quantity $\mathfrak{S}$ defined as
\begin{equation}
\mathfrak{S}(t, \omega, k)=\int_0^{\infty} \int_{-\infty}^{\infty} \psi(x, t-t^{'})e^{- t^{'} /T_{av}} e^{-ikx + i\omega t^{'}}dx dt^{'}.
\end{equation}
This quantity can be interpreted as a spectrum measured by a spectrometer with the resolution $\Delta \omega = \frac{1}{T_{av}}$.  Obviously, to resolve the newly generated frequency from the frequency of the initial polariton state the time $T_{av}$ must be large enough. If $T_{av} \rightarrow \infty$ then $\mathfrak{S}$ is simply a Fourier representation of the field. Let us also remark that, contrary to Fourier transform, the quantity $\mathfrak{S}$ varies on time. This allows to watch how the new frequencies are being generated during the development of the instability 

\begin{figure}[h]
\includegraphics[width=\linewidth]{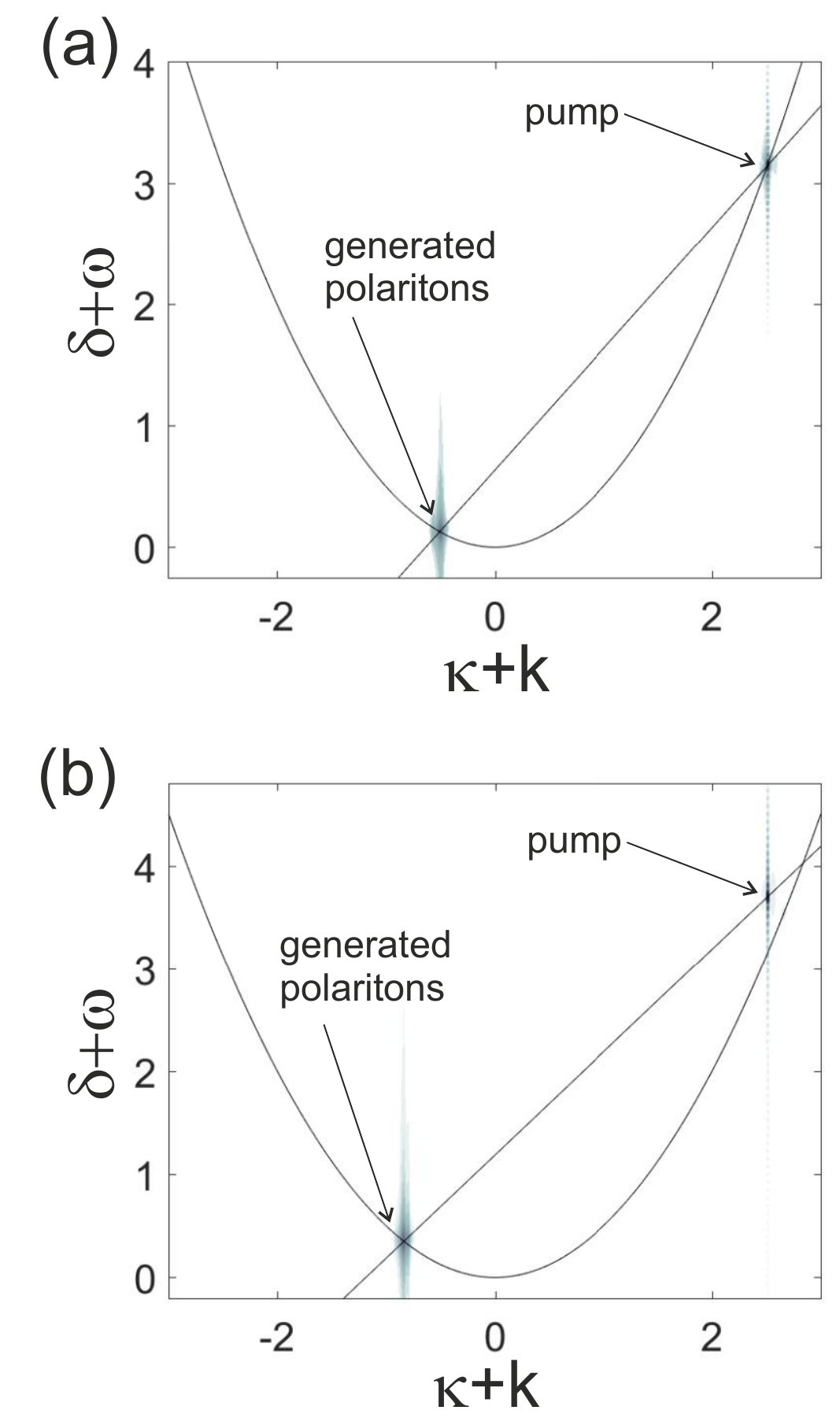}
\caption{The spectral representation of the polariton fields obtained in numerical modeling. Panel (a) corresponds to the zero detuning and panel (b) to the non-zero detuning of the pump. The curved lines show the dispersion of the linear polaritons, the straight lines show the appropriately shifted dispersion of the sound responsible for the instability.}
\label{fig:3}
\end{figure}

The calculated spectrum $\mathfrak{S}$ is shown in  Fig.~\ref{fig:3}. To illustrate the linear stage of the instability the spectrum $\mathfrak{S}$ is shown for the time when the spectral patterns produced by the instability is already well visible, but the generated field does not modify the polariton state much (the generated excitations can be considered as linear perturbations and thus the developed linear theory is applicable). The developed stage of the instability is probably of more interest from the experimental point of view and this problem is also addressed in the paper.

\begin{figure}[h]
\includegraphics[width=\linewidth]{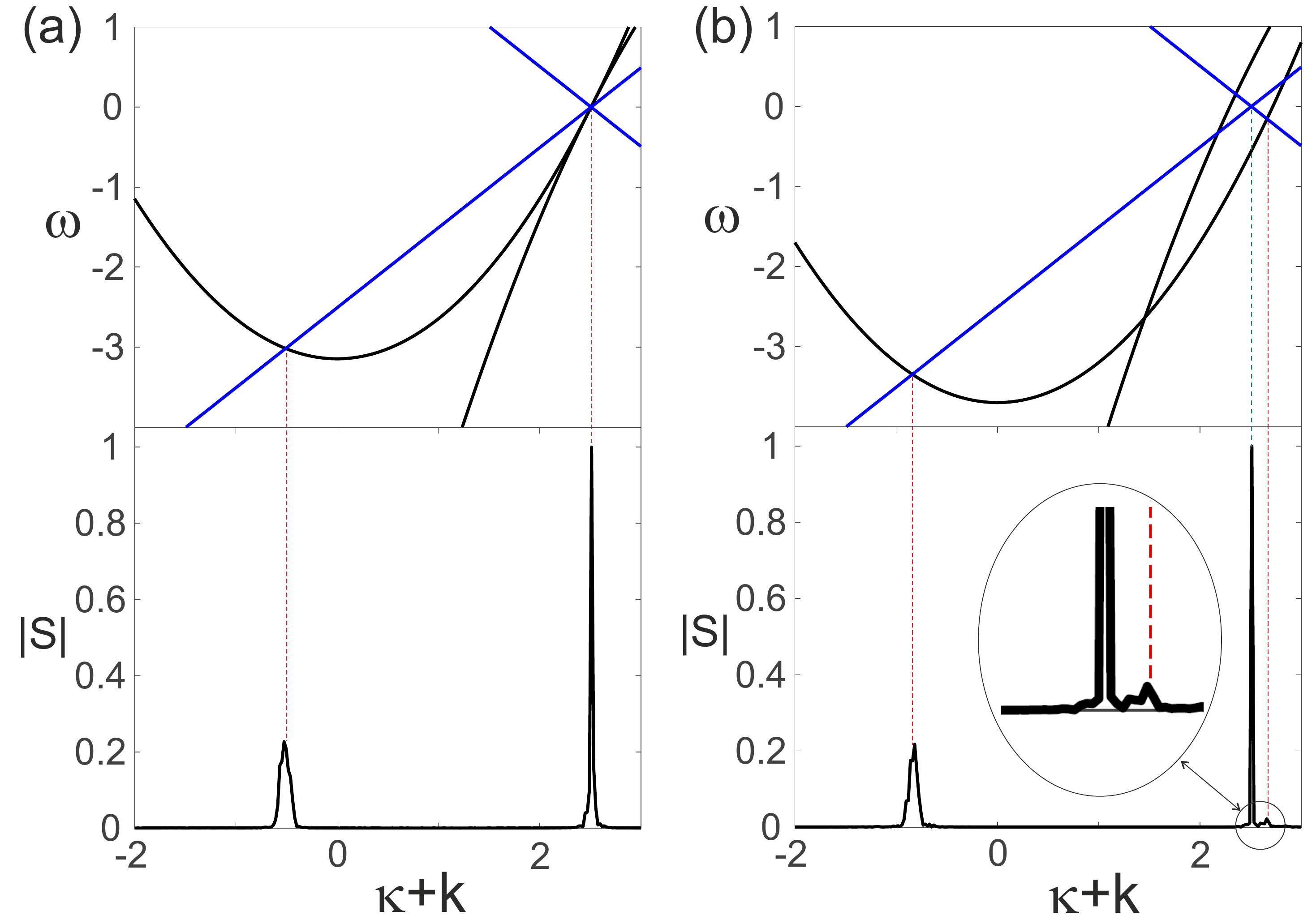}
\caption{(a) The dispersion relation of the weak excitations  and the spatial spectrum of the computed polariton field for the case of the resonant pump. The dispersion characteristics are the same as in Fig.~\ref{fig:3} but shifted by the wavevector of the pump. The red lines are marking the positions of the resonances generating the instability. (b) The same as in (a) but for the pump with a detuning $\mu=3.7$. }
\label{fig:4}
\end{figure}

To find out if the developed linear theory predicts the positions of the resonances we compared the spatial characteristics of the polariton order parameter function  with the resonances given by the crossing of the dispersion characteristics. Let us note that the total field containing the background and the growing excitation is not spatially uniform anymore and its instructive to calculate its spatial spectrum defined as $S(k)=\int \psi(x) \exp(-i k x) dx$. In experiments this quantity can be extracted from directly measured radiation intensity in the far-field zone and under some assumptions is simply the far-field spectrum.  In Fig.~\ref{fig:4} it is seen that the positions of the generated spectral lines are predicted very accurately by the resonant condition.

It is also possible to pump the system out of the resonance. Then the linear excitations on the intense polariton state have different dispersion, see Fig.~\ref{fig:1}(b).
It is seen that the shift of the dispersion characteristics of the polaritons causes two additional resonances.

The real and the imaginary parts of the frequencies of the linear excitations in the vicinity of the dispersion crossing points are shown in Fig.~\ref{fig:2}(a).
The analysis shows that one of the two additional resonances causes an instability.
The spectral representation of the numerically calculated field is shown in Fig.~\ref{fig:3}, where it is seen that the pump is shifted off the dispersion curve of the linear polaritons.
The reason why only one resonance is seen is explained by the fact that the additional resonance leads to the  instability with much smaller growth rate than the growth rate of the instability generated by the first resonance.

The normalized spectrum of the calculated polariton field is shown in Fig.~\ref{fig:4}(b).
There it is possible to identify the spectral peaks corresponding to the pump and to the polaritons produced due to the instability. It is worth mentioning that the peak corresponding to the second resonance is weak but clearly visible in the inset. One can see that the position of the spectral lines observed in direct numerical simulations fits to the predictions of the resonant condition very well. 

\begin{figure}[h]
\includegraphics[width=\linewidth]{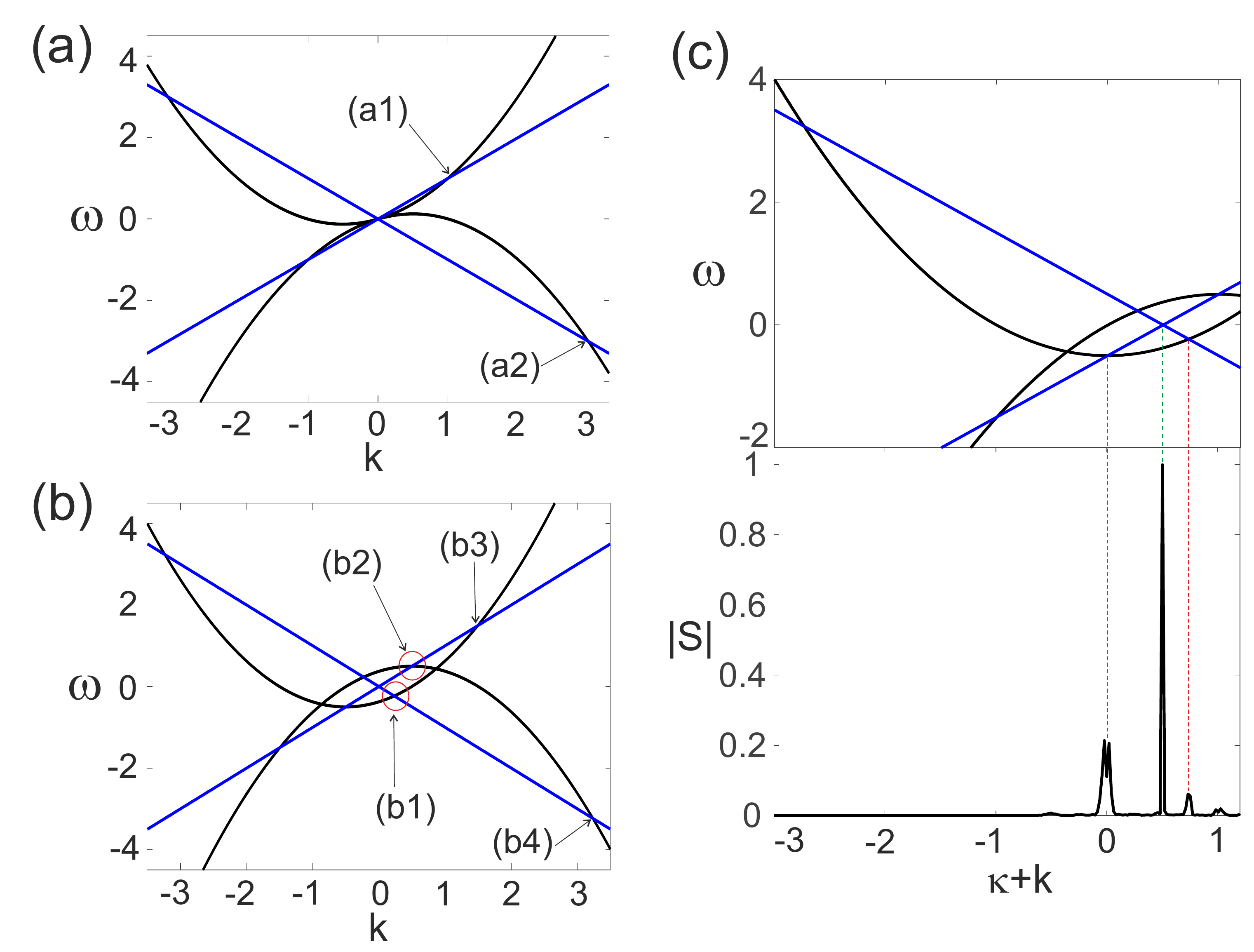}
\caption{The same as in Fig.~\ref{fig:1} but $\kappa=0.5024$ The frequency of the resonant pump is $\delta=0.1262$ (panel (a)) and the frequency detuning of the resonant pump is $\mu=0.5$ (panel (b)). (c) The dispersion relation of the weak excitations  and the spatial spectrum of the computed polariton field for the case of the pump with a non-zero detunig. The dispersion characteristics are the same as in (b) but shifted by the wavevector of the pump. The red lines are marking the positions of the resonances generating the instability.}
\label{fig:7}
\end{figure}

Now let us consider the case when the polaritons directly generated by the coherent pump have phase velocity smaller than the velocity of the sound.
The dispersion characteristics of the linear excitations are shown in  Fig.~\ref{fig:7} for the zero and non-zero detunings of the pump.  

\begin{figure}
\includegraphics[width=\linewidth]{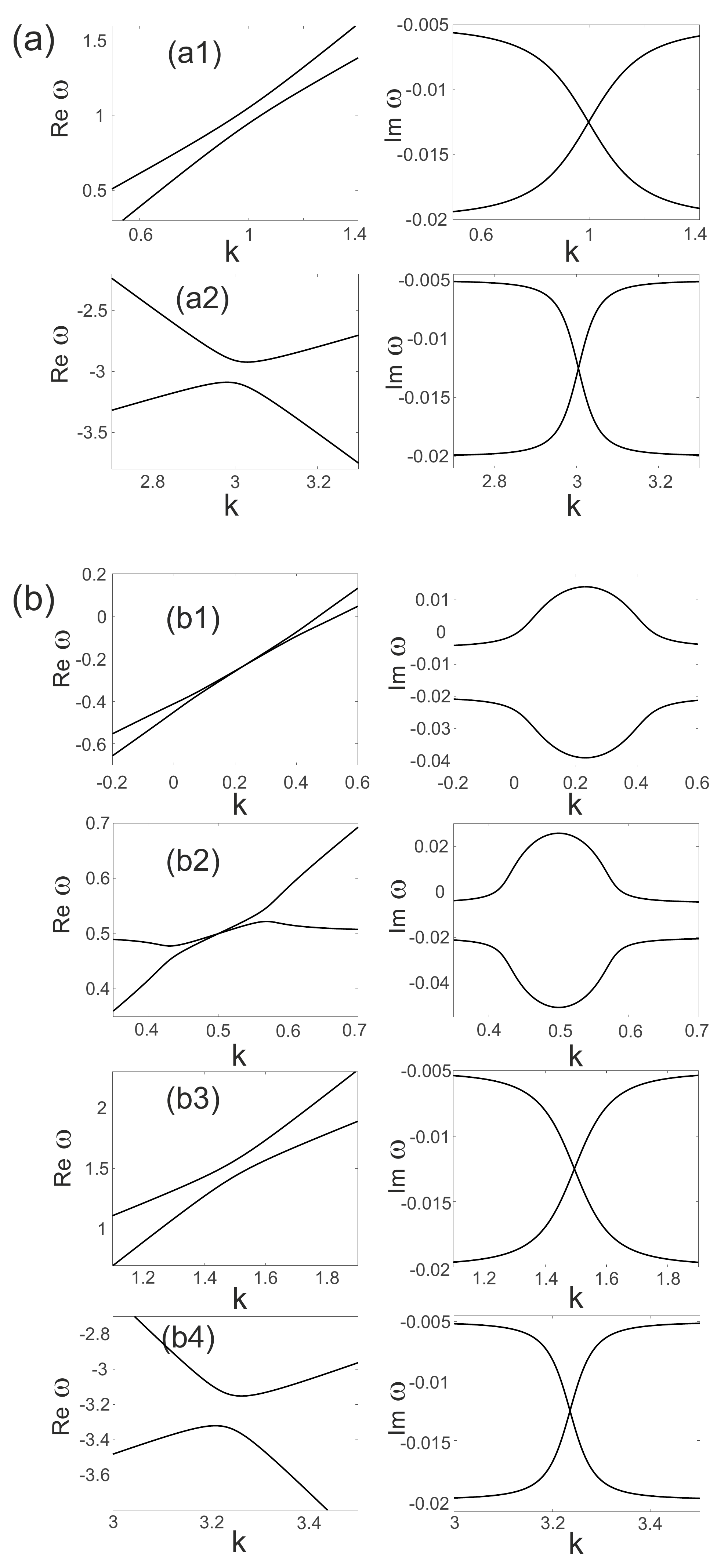}
\caption{(a) The dependencies of the real (left column) and imaginary (right column) parts of the frequencies of the weak excitations as functions of their wavevectors. The row marked as (a1) corresponds to the crossing marked as (a1) in Fig.~\ref{fig:7}(a), the row marked as (a2) corresponds to the crossing marked as (a2) in Fig.~\ref{fig:7}(a). The population of the unperturbed polariton state is $N_0=1.0$. (b) The same as in (a) but for the pump with $\mu=0.5$. }
\label{fig:8}
\end{figure}

The analysis of the dispersion characteristics in the vicinity of the crossing points reveals that in the case of the resonant pumping no instabilities appear, both existing resonances lead to the hybridization of the modes but not to the sound generation, see  Fig.~\ref{fig:8}.
The formation of the spatially uniform polariton state was observed in numerical simulation when the amplitude of the pump was $P=0.02$ and it confirmed that no instability is generated in this case.

If the pump is detuned then two additional resonances at low $k$ appear and both these resonances generate sound waves because of the phonon-polariton instability.
The spectra of the linear excitations in the vicinity of the resonances is shown in Fig.~\ref{fig:8}.

To check the analytic results we performed numerical simulation with pump amplitude $P=0.3743$ needed to obtain the polariton state with the population $N_0=1$.
First the interaction between the polaritons and the acoustic waves was set to zero. When the stationary state formed it was perturbed by weak noise in both components and the interaction between the polaritons and the sound was made finite $g=0.075$. 

\begin{figure}[h]
\includegraphics[width=\linewidth]{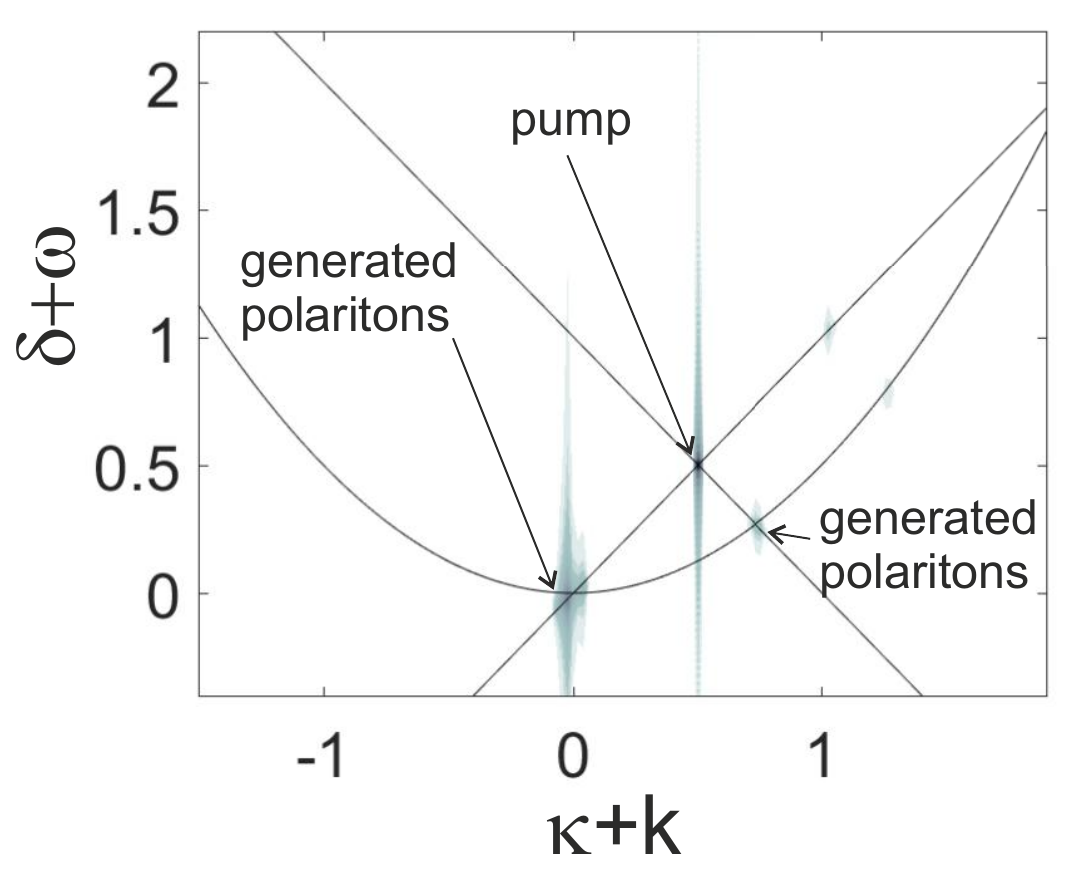}
\caption{The spectral representation of the polariton fields obtained in numerical modeling. Thin curve shows the dispersion of the linear polaritons, the straight lines show the appropriately shifted dispersion of the sound responsible for the instability.}
\label{fig:10}
\end{figure}

The spectral representation of the polariton field is shown in Fig.~\ref{fig:10}.
The spectral patterns corresponding to both resonances are seen in the spectral diagram.
The spectrogram corresponds to the quite developed instability and so the additional spectral patterns corresponding to three-wave mixing is also noticeable in the diagram.   

The comparison of the analytics and the numerics is illustrated in Fig.~\ref{fig:7}(c) showing the calculated spatial spectrum of the polariton field.
It is seen that the positions of the generated spectral lines coincide with the linear resonances exactly.

\subsection{Properties of linear excitations in the presence of nonlinear polariton-polariton interactions}

\begin{figure}[h]
\includegraphics[width=\linewidth]{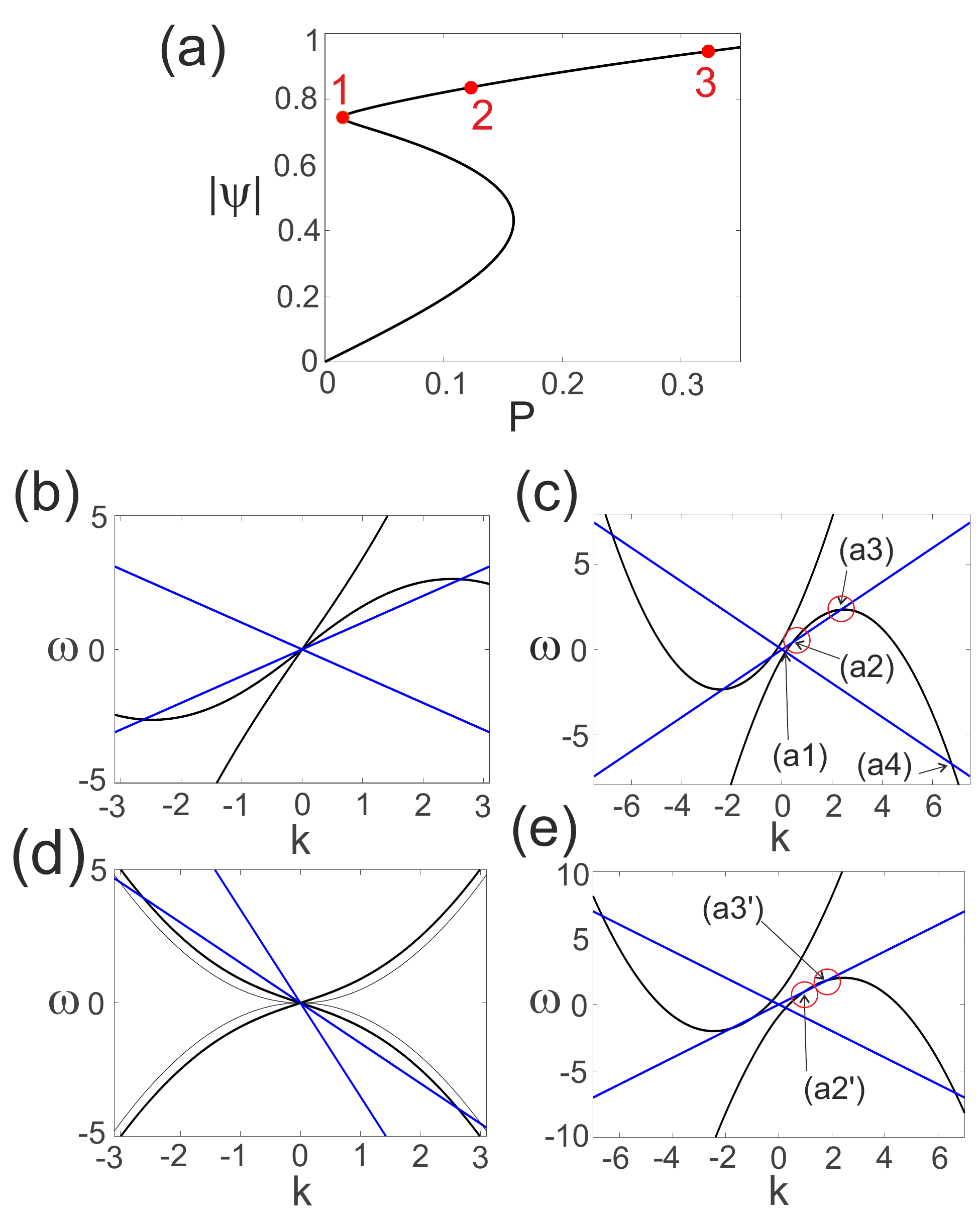}
\caption{Panel (a) The bifurcation diagram for $\delta=3.7$ and $\alpha=1$. Panels (b) and (d) The spectrum of linear excitations on the background corresponding to the point marked as '1' in (a). The pump $P=0.01489$ produces the polariton state with the amplitude $|\psi|=0.744$. The black curves correspond to the dispersion of the linear polariton excitations, the blue ones - to the phonon excitations. Panel (b) shows the spectrum in the laboratory reference frame, the panel (d) - in the reference frame moving with the velocity $v_{rf}=\kappa$. Thinner line in panel (d) shows the dispersion of the polaritons not interacting with each other. Panels (c) and (e) The dispersion characteristics for the weak excitation on the states marked as '2' (panel (c)) and '3' (panel (e)) in  Fig.~\ref{fig:12}. The red circles mark the crossing resulting in the instability for sufficiently strong coupling between the phonons and the polaritons. Black curves correspond to the dispersion of the polaritons and the blue curves - to the dispersion of the phonons.}
\label{fig:12}
\end{figure}

Now it is logical to consider the problem with nonlinear polariton-polariton interaction.
The case of interest is defocusing nonlinearity corresponding to repelling inter-polariton interactions. Discussing the states without the acoustic component we note that depending on the pump power the states can exhibit bistability.
The bifurcation diagram corresponding to this case is shown in Fig.~\ref{fig:12}. The upper branch of the bifurcation diagram is stable, the lower is partially stable and the intermediate is unstable.

Let us consider the weak excitation on the background belonging to the upper branch of the bifurcation diagram.
As before we first study the case $g=0$ when polaritons do not interact with phonons.
At the folding point of the bifurcation diagram marked as '1' in  Fig.~\ref{fig:12} the spectrum of the polariton excitations is Bogoliubov one, see Fig.~\ref{fig:12}(b and d).
To make it even more clear we plotted the spectrum in the reference frame moving with the velocity $v_{rf}=\kappa$ ($\kappa$ is the wavevector of the pump).
For reference the thinner line shows the dispersion of the linear excitations without polariton-polariton interactions.  

Let us consider the polariton states belonging to the upper branch of the bifurcation diagram.
The spectrum of linear excitations for the points marked as '2' and '3' in Fig.~\ref{fig:12} is shown in Fig.~\ref{fig:12}(c and e).

\begin{figure}[h]
\includegraphics[width=\linewidth]{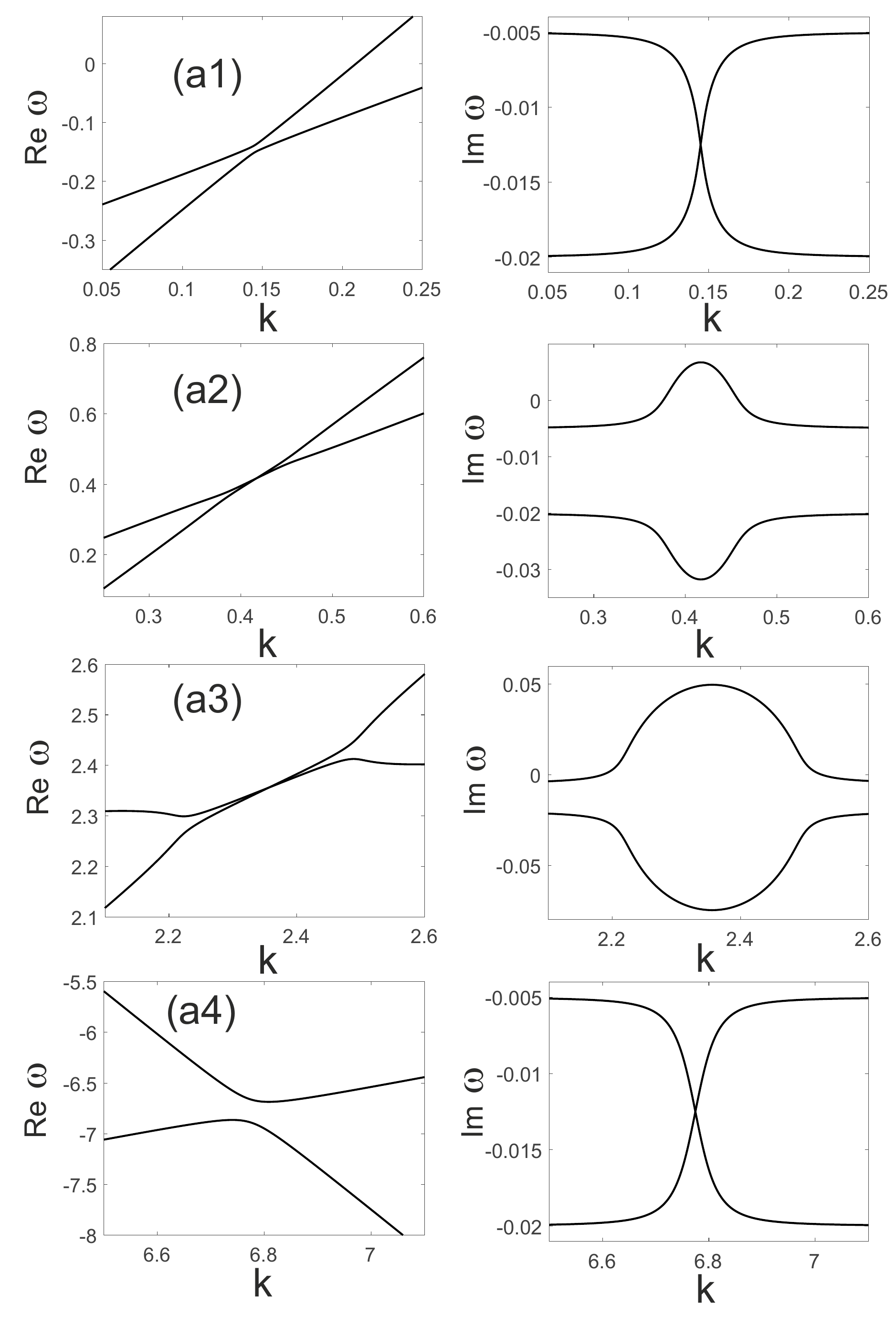}
\caption{The dispersion characteristics in the vicinity of the crossings marked as 'a1'-'a4' in Fig.~\ref{fig:12}(c). The polariton-phonon interaction strength is $g=0.075$.}
\label{fig:15}
\end{figure}

It is seen that the tangential point of the polariton dispersion characteristics disappears.
This results in two additional crossings of the dispersion of the polaritons and the phonons, these points are marked as 'a1' and 'a2'. The resonances causing the instability are marked by red circles in Fig.~\ref{fig:12}(c).
The complete information about the dispersion of the linear excitations in the vicinity of the resonances is summarized in Fig.~\ref{fig:15}.

\begin{figure}
\includegraphics[width=\linewidth]{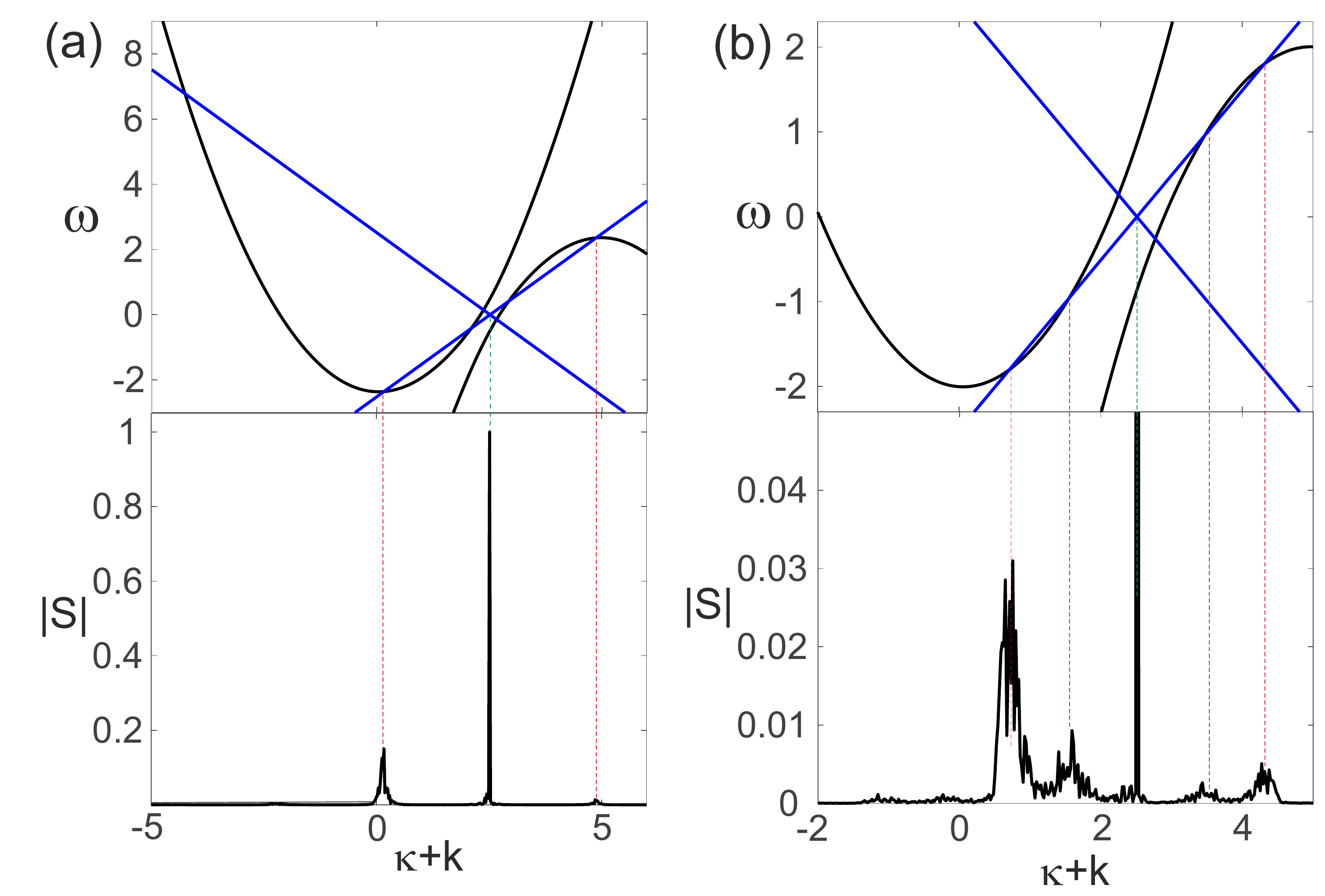}
\caption{(a)The dispersion characteristics of the polaritons and the phonons and the spatial spectrum of the  polariton field generated by the instability of the state '2' in  Fig.~\ref{fig:12}. (b)The dispersion characteristics of the polaritons and the phonons and the spatial spectrum of the  polariton field generated by the instability of the state '3' in  Fig.~\ref{fig:12}. The pump $P=0.3226$ produces the polariton state with amplitude $|\psi|=0.9458$.}
\label{fig:17}
\end{figure}

To check if the wavevector of the forming polariton state is predicted by the instability condition we compared the results of the numerical simulations against the predictions of the analytic consideration.
As one can see in Fig.~\ref{fig:17} the position of the polariton spectral line is predicted by the analytics very precisely.
The excitation efficiency for the second resonance generating the instability is much smaller than that for the first resonance and so the spectral line corresponding to the second resonance is not visible.

It is worth mentioning that if the pump is increased then the resonances 'a2' and 'a3' go towards each other, compare the resonances 'a2' and 'a3' in panel (c) of Fig.~\ref{fig:12} with the resonances 'a2$^{'}$' and 'a3$^{'}$' in panel (e).
That is why at higher pump intensities both resonances produce new polariton states of comparable intensities, see the spectrum of the polariton field generated by the resonances 'a2$^{'}$' and 'a3$^{'}$' in  Fig.~\ref{fig:17}(b).
If the pump is increased even further then the resonances collide and disappear.
So at high levels of pump the nonlinear polariton-polariton interaction can stabilize pure polariton states.
In the absence of the interaction this stabilization does not happen.

\subsection{Stationary hybrid states appearing at the nonlinear stage of phonon-polariton instability}

Now let us consider the stationary states forming from the developing instability.
We start with the case without polariton-polariton interaction. The numerical experiment was done as follows.
We set the low intensity of the pump and calculated until stationary state formed.
Then we measure the stationary amplitude of the polariton state with the wavevector around the wavevector of the pump.

The amplitude of the directly excited field can be defined as  $a_1=\sqrt{ \int_{k_p-\Delta k}^{k_p+\Delta k} |\psi(k)|^2 dk/L}$. If the components are well resolved spectrally then $a_1$ does not depend on the integration limits provided that $\Delta k$ is much bigger than the with of the spectral line of the component. If the component is absolutely uniform spatially  then the introduced amplitude $a_1$ is equal to the real amplitude of the state. We can also introduce $a_2=\sqrt{ \int_{-\infty}^{\infty} |\psi(k)|^2 dk/L - a_1^2}$. This quantity is equal to the amplitude of the second component provided that there are only two components and that they are spatially uniform. If this is not so then $a_2$ can be considered as a measure of the portion of the polaritons that are not excited directly by the coherent pump.

\begin{figure}
\includegraphics[width=\linewidth]{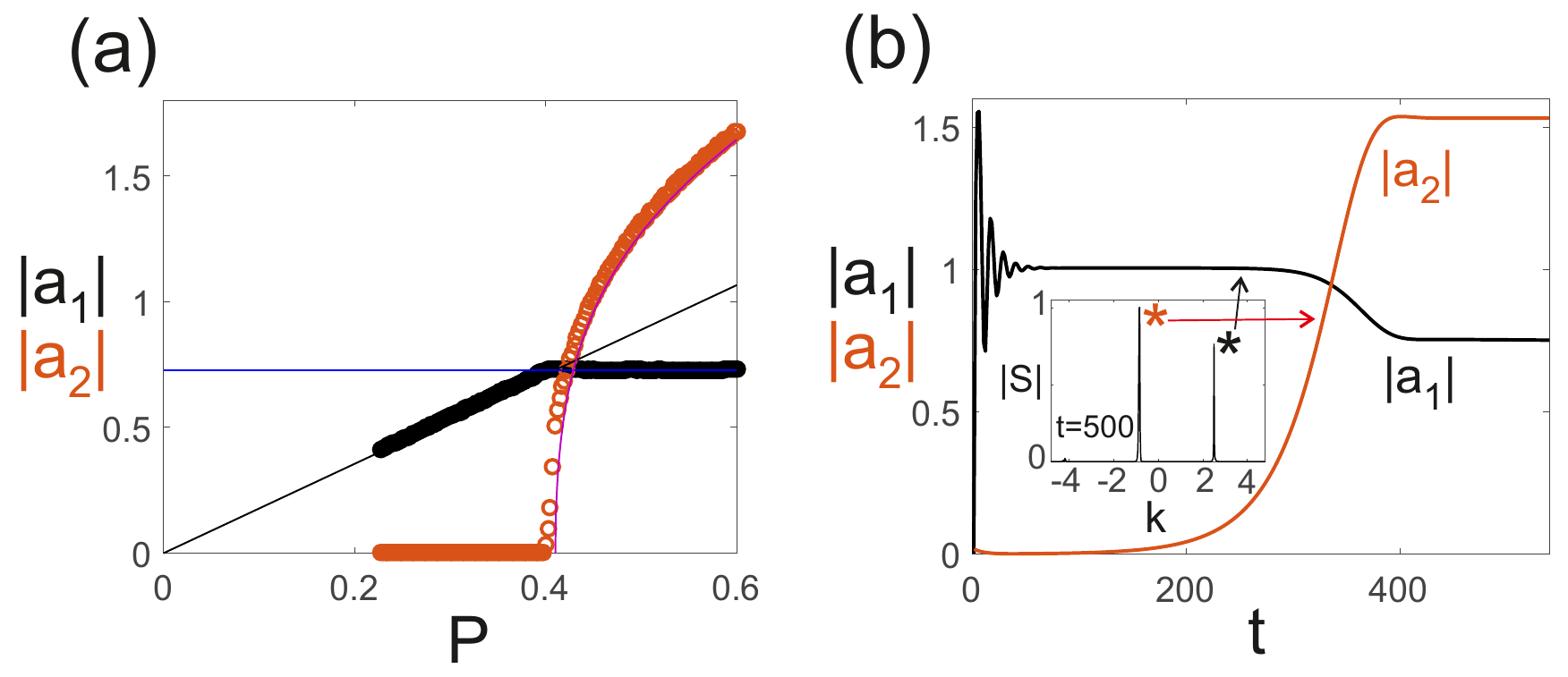}
\caption{(a) The dependencies of the amplitudes of the directly pumped $|a_1|$ and parametrically generated $|a_2|$ states on the amplitude of the pump. The black and red circles show the numerically calculated values. The dependency of the amplitude of the polariton state on the pump in the absence of the coupling of the phonons and polaritons is shown by thin black line. The dashed blue line marks the amplitude of the directly excited state found in the framework of the developed perturbation theory allowing to calculate stationary amplitudes. The analytically found dependency of the amplitude of parametrically excited polariton state on the pump is shown by solid magenta curve. (b) The temporal evolution of the amplitudes for $P=0.554$. The spatial spectrum of the polariton field at $t=500$ is shown in the inset.}
\label{fig:19}
\end{figure}

After reaching a stationary state and measuring of its amplitude we perturb the state by weak noise and increase the pump by a small increment. After reaching a new stationary state the amplitudes are measured.
This way we can find the dependence of the amplitude of the states as functions of the pump intensity. This dependencies are shown in Fig.~\ref{fig:19}. 

One can see that when the instability threshold is exceeded and the phonons start being generated the amplitude of the polariton state with the wave vector equal to the wave vector of the pump does not grow, in the same time the amplitude of the generated polariton state growth with the intensity of the pump. In a certain sense one can say the after the instability threshold the energy of the pump goes not to the polariton directly but through the parametric process to the acoustic mode and to the new polariton state. The temporal dynamics of the amplitudes is shown in Fig.~\ref{fig:19}(b). In the simulations of the dynamics we took weak noise as initial conditions.

Using formulas (\ref{a1})-(\ref{a2+noKerr}) it is possible to calculate the amplitudes of the generated polariton states and to compare them with the numerically obtained curves. The analytically and the numerically obtained curves are shown in Fig.~\ref{fig:19} and the comparison of the dependencies  shows that they fit each other perfectly.

The dynamics becomes much richer in the presence of nonlinear polariton-polariton interactions  In particular complex spatiotemporal evolution and the effect of hysteresis have been observed in numerical simulations. Let us describe the scheme of the numerical experiment. The initial conditions were chosen in a form of weak noise in both the polariton and the acoustic components. The initial pump was of low intensity. The calculation have been performed for the time much longer compared to all characteristic times of the system. Then the spatial spectrum was recorded and the amplitudes of the directly and parametrically excited states were extracted from the numerical data. 

Then the stationary distributions of field in the  previous simulation was perturbed by weak noise and taken as the initial conditions for the new round of the numerical simulation with increased amplitude of the pump. This way the spatial spectrum and the amplitudes were measured for the increased pump. Repeating this procedure we measured the dependencies of the spatial spectrum and the amplitudes of directly and parametrically excited states on the pump amplitudes. 

\begin{figure}
\includegraphics[width=\linewidth]{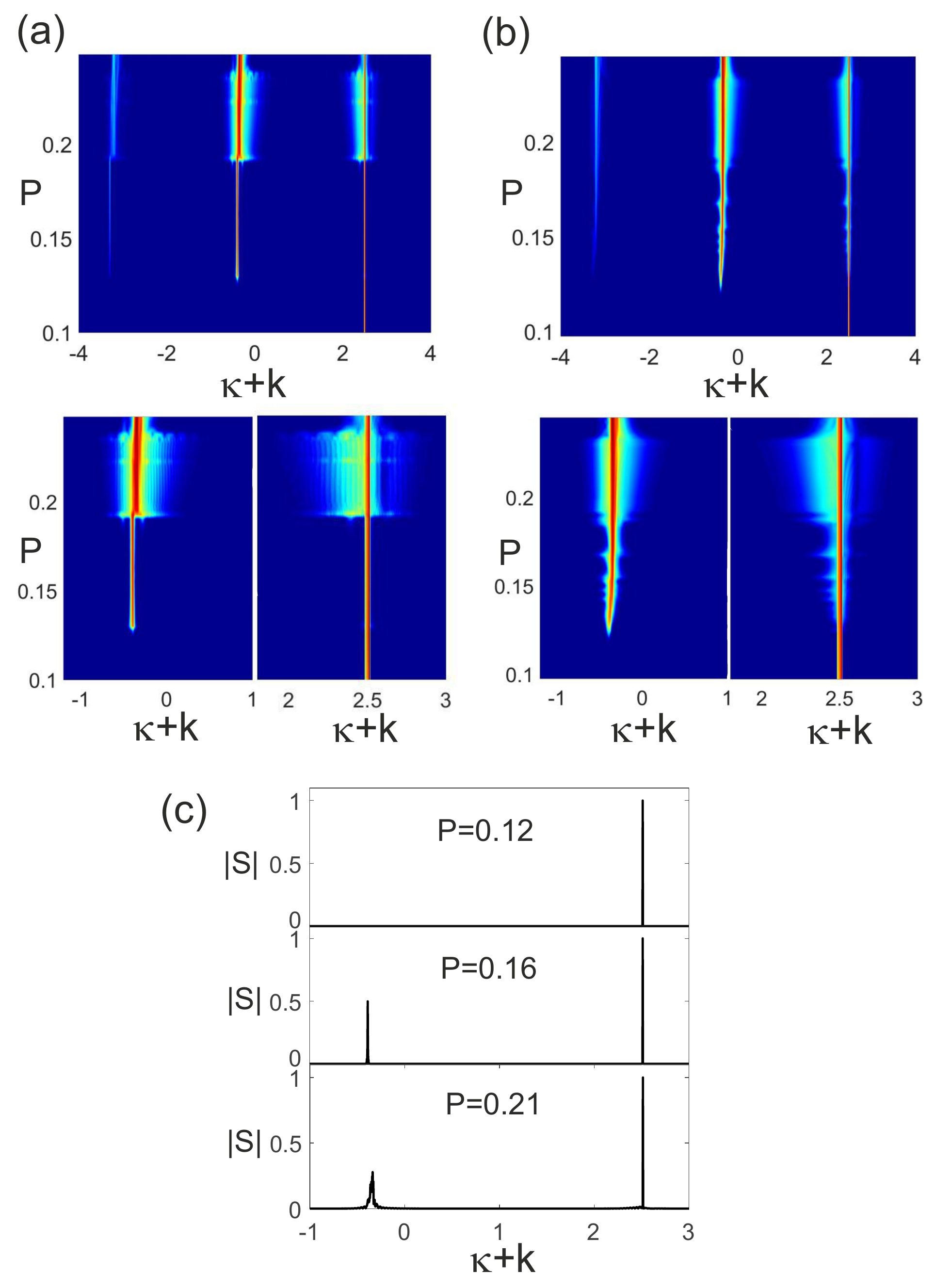}
\caption{The numerically calculated dependency of the spatial spectrum of polariton stationary states on the amplitude of the pump. Panel (a) shows the variation of the pump from lower to higher values. The variation of the spatial spectrum when the pump is decreasing is shown in panel (b). The spatial spectra of the stationary polaritons are shown in panel (c).  The parameters are $\kappa=2.51$, $\delta=3.1$, $g=0.75$, $\alpha=0.1$.}
\label{fig:20}
\end{figure}

The dependence of the spatial spectrum on the pump amplitude is shown in  Fig.~\ref{fig:20}(a). At lo intensities of the pump the spectrum contains only one spectral line corresponding to the directly excited state. Then at some threshold pump a second spectral line corresponding to parametrically excited state appears. It is good to notice that the position of the spectral line corresponding to the generated state depends on the pump in the presence of nonlinear polariton-polariton interactions. It is is interesting that if we continue to increase the pump then there is a second threshold when the spectral lines of the polaritons get broadened, see Fig.~\ref{fig:20}(c) showing spatial spectra for different pump intensities. 

\begin{figure}
\includegraphics[width=\linewidth]{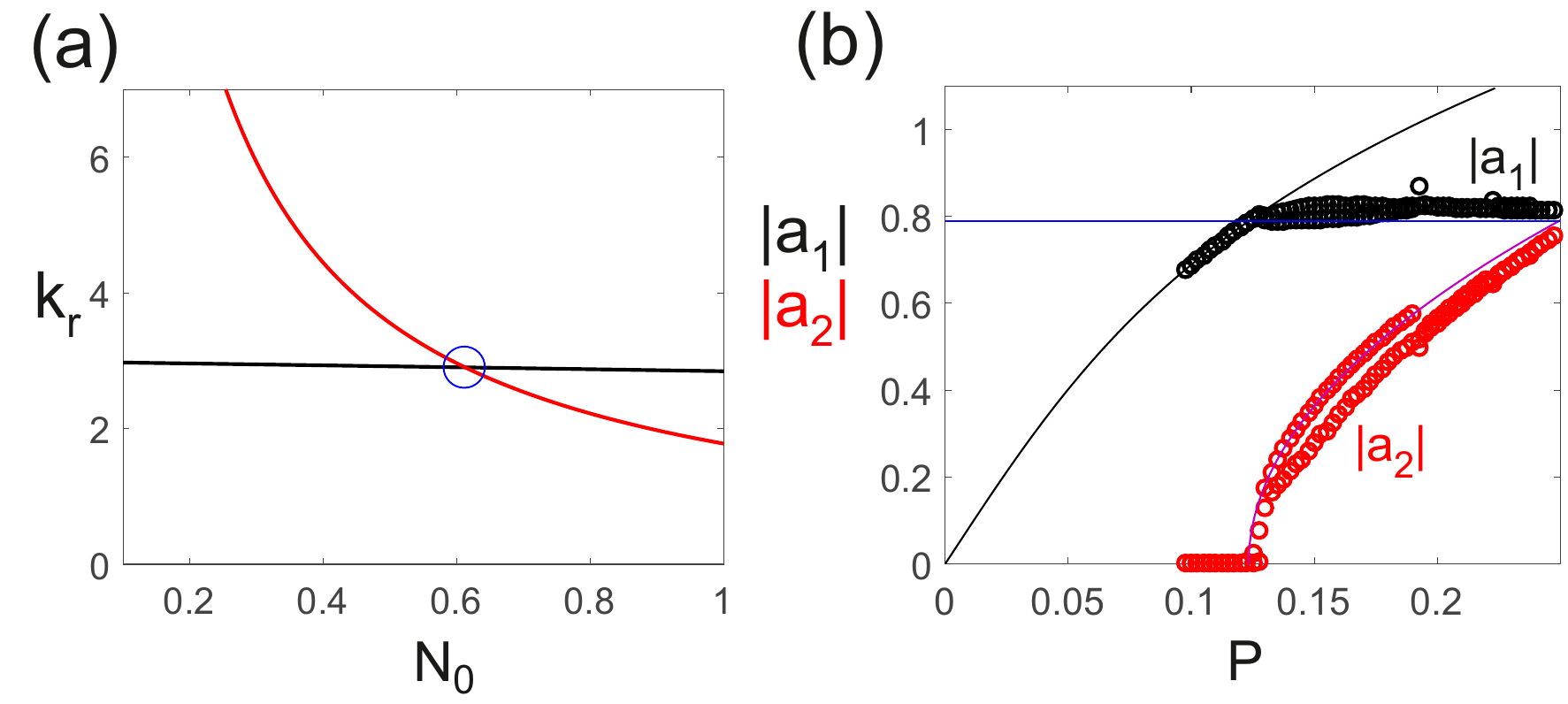}
\caption{Panel (a) shows the numerical solution of the equation defining the population of the polariton state and the wavevector of the phonons at the instability threshold. The dependencies of the amplitudes of polariton stationary states on the amplitude of the pump are shown in panel (b) The thin black curve show the dependency of the polariton state on the pump in the absence of the acoustic field. The thin blue line marks the analytically calculated amplitude of the polariton state at the instability threshold. The thin magenta line shows the analytically found dependency of the amplitude of the parametrically excited polaritons on the amplitude of the pump. The black and the red circles correspond to the amplitudes extracted from the direct numerical simulations. The parameters are $\kappa=2.51$, $\delta=3.1$, $g=0.75$, $\alpha=0.1$.}
\label{fig:21}
\end{figure}

A natural question is what happens if the amplitude of the pump goes back to smaller values. The numerical simulations show that the forward and the reverse paths are not identical, compare panels (a) and (b) of Fig.~\ref{fig:20}. It is also seen if we compare the dependencies of the amplitudes of the directly and parametrically excited polaritons on the pump amplitudes, see Fig.~\ref{fig:21}. So we can conclude that the the hybrid acousto-polariton states can be different at the same pump amplitude depending on how the state was created.

A simple theory developed in the previous section can give us the values of the directly and parametrically excited components in the vicinity of the instability threshold. Then we assume that the amplitude of the parametrically excited state is small and thus does not affect the position of the resonance. However the amplitude of the directly excited state affects the resonance position and this has to be taken into account. Formula (\ref{a1}) shows how the critical amplitude of the directly excited state $a_1$ (at which the instability sets in) depends on the resonant wave vector. At the same time the value of the resonant wavevector depends on $a_1$. Solving these two equations one can find the critical amplitude $a_1$  and the resonant wave vector $k_r$. Graphical solution is presented in  Fig.~\ref{fig:21}(a). Then using formula (\ref{a2}) it is possible to find the amplitude of the parametrically excited state. The analytically calculated dependency of the amplitude of the parametrically excited state on the pump is shown in Fig.~\ref{fig:21}(b). At low pump intensity the perturbation theory matches the results of numerical simulations well, but at higher pumps it fails to describe the states observed in numerical simulations. 

\section{Conclusion}

In the present paper the mutual dynamics of polaritons and acoustic waves is considered in mean field approximation. It is shown that there may be resonant interaction between the acoustic waves and the polaritons. The resonant condition was found and it is demonstrated that polaritons moving with the velocity greater than the velocity of the sound can excite coherent acoustic waves and new polaritons with a lower frequency. This process is analogous to the parametric processes used, for example, in optical parametric oscillators for new frequencies generation.  At the same time subsonic polaritons do not exhibit instability but the spectrum of linear excitations is of hybrid nature with the polaritonic and acoustic components. 

The theory is extended for the case of coherently driven systems with finite losses in both the polariton and the acoustic subsystems. The polariton states were found and the spectra of linear excitations are analyzed. It is demonstrated that the pump changes the positions of the resonances and can lead to the appearance additional unstable modes. In the presence of nonlinear polariton-polariton interactions the elementary excitations are combined particles consisting of polariton excitations of Bogoliubov kind and the phonons. The polariton-polariton interaction affects the stability of the system. At the pump strong enough different resonance can collide and disappear, Thus the nonlinear polariton-polariton interaction can stabilize the system and suppress the generation of phonons.

The formation of the stationary states was analyzed by direct numerical simulations and by a perturbation theory. It is demonstrated that the developed theory describes the stationary states well in the vicinity of of the instability threshold. 

The collective dynamics of the polariton and phonons becomes complicated in the presence of nonlinear polariton-polariton interactions if the pump intensity is strong enough. It was shown by direct numerical simulations that the system can exhibit such effects as multistability and hysteresis when the stationary state depends not only on the pump but on the evolution of the system. It is worth mentioning that at high pump intensities the spatial spectrum of the polaritons broadens which means that long waves are excited in the system.

The effects discussed in the paper can possibly have not only fundamental but also practical importance. For example this can be used for the development of acousto-polariton lasers or for the downconversion of the polariton frequencies.
\section{Acknowledgements}

 This work was supported by mega-grant No. 14.Y26.31.0015 of the Ministry of Education and Science of the Russian Federation. AVY acknowledges support from the Government of the Russian Federation (Grant 074-U01) through the ITMO University fellowship. AVN acknowledges support from RFBR Grant No. 18-32-00434. IAS acknowledges support from goszadanie no 3.2614.2017/4.6 of the Ministry of Education and Science of the Russian Federation.

%\clearpage
\bibliographystyle{apsrev4-1}
\bibliography{references}
\end{document}